\documentclass[aps,prd,twocolumn,unsortedaddress,showpacs,nofootinbib]{revtex4-1}

\usepackage{amsmath,amssymb,graphicx}
\usepackage{color}
\usepackage{pslatex}
\usepackage{pdfpages}
\usepackage{graphicx}
\usepackage{psfrag}

\usepackage{graphicx}
\usepackage{hyperref}

\usepackage[normalem]{ulem}

\begin{document}

\title{Fitting the Fermi-LAT GeV excess: On the importance of including the propagation of electrons from dark matter}

\author{Thomas Lacroix}
\affiliation{UMR7095, Institut d'Astrophysique de Paris, 98 bis boulevard Arago, 75014 Paris, France}
\email{lacroix@iap.fr}
\author{C\'{e}line B\oe hm}
\affiliation{Institute for Particle Physics Phenomenology, Durham University, Durham, DH1 3LE, United Kingdom}
\affiliation{LAPTH, Universit\'{e} de Savoie, CNRS, BP 110, 74941 Annecy-Le-Vieux, France}
\email{c.m.boehm@durham.ac.uk}
\author{Joseph Silk}
\affiliation{UMR7095, Institut d'Astrophysique de Paris, 98 bis boulevard Arago, 75014 Paris, France}
\affiliation{The Johns Hopkins University, Department of Physics and Astronomy,
3400 N. Charles Street, Baltimore, Maryland 21218, USA}
\affiliation{Beecroft Institute of Particle Astrophysics and Cosmology, Department of Physics,
University of Oxford, Denys Wilkinson Building, 1 Keble Road, Oxford OX1 3RH, United Kingdom}
\email{silk@iap.fr}

\date{\today}

\begin{abstract}
An excess of gamma rays at GeV energies has been pointed out in the Fermi-LAT data. This signal comes from a narrow region centred around the Galactic center and has been interpreted as possible evidence for light dark matter particles annihilating either into a mixture of leptons-antileptons and $b\bar{b}$ or into $b \bar{b}$ only. Focusing on the prompt gamma-ray emission, previous works found that the best fit to the data corresponds to annihilations proceeding predominantly into $b\bar{b}$. However, here we show that omitting the photon emission originating from primary and secondary electrons produced in dark matter annihilations, and undergoing diffusion through the Galactic magnetic field, can actually lead to the wrong conclusion. Accounting for this emission, we find that not only are annihilations of $\sim 10\ \rm GeV$ particles into a purely leptonic final state allowed, but the democratic scenario actually provides a better fit to the spectrum of the excess than the pure $b \bar{b}$ channel.  We conclude our work with a discussion on constraints on these leptophilic scenarios based on the AMS data and the morphology of the excess.
\end{abstract}


\maketitle

\section{Introduction}
After several decades of remarkable experimental development, evidence for dark matter (DM) particles still remains to be found. One important technique that has made dramatic progress in the last few years is indirect detection, which aims to detect the annihilation or decay products of DM particles in dense environments such as the central region of our Milky Way halo. In particular, the recent gamma-ray data from the Fermi-LAT (Large Area Telescope) experiment has enabled the community to constrain the thermal DM paradigm and set important bounds on the DM self-annihilation cross section as a function of the DM mass, for various final states (see, e.g., Refs.~\cite{Fermi_extragalactic,Fermi_dwarfs}).

However, a few years ago, the possibility of a gamma-ray excess at low energies (between 1 and 10 GeV), in a narrow region around the Galactic center (GC)---smaller than $10^{\circ} \times 10^{\circ}$ \cite{Vitale_Fermi}---led several authors to speculate that this could be a manifestation of DM annihilations into either a mixture of $b \bar{b}$ and leptons-antileptons final states, or $b \bar{b}$ final states only \cite{Hooper,Hooper_Goodenough_excess,GordonMacias,Abazajian_GeV_excess,Daylan_GeV_excess}. While this excess could be attributed to astrophysical sources---like the central point source \cite{Ruchayskiy_Fermi_excess}, a burst injection of electrons \cite{electron_burst_excess}, a population of cosmic-ray protons \cite{protons_Profumo_excess}, or unresolved millisecond pulsars (see, e.g., Ref.~\cite{GordonMacias})---a DM interpretation is nevertheless possible. In the case of a pure $b \bar{b}$ final state, a DM mass of 30 GeV would be favored, while the DM mass should be about 23.5 GeV if the final state contains 45$\%$ leptons and 55$\%$ $b$ quarks \cite{GordonMacias}. In Refs.~\cite{Hooper,GordonMacias} it was also found that a DM mass of 10 GeV is required if the final state contains 90$\%$ leptons and 10 $\%$ $b$ quarks but the quality of the fit was better for the $b \bar{b}$ channel, thus leading the authors to prefer a large fraction of $b$ quarks in the final state. Note that throughout this paper, the term ``leptons" refers to democratic annihilation into leptons, i.e., a combination of the $e^{+}e^{-}$, $\mu^{+}\mu^{-}$, $\tau^{+}\tau^{-}$ final states, with 1/3 of the annihilations into each of these channels.

These conclusions were obtained by only taking into account the prompt gamma-ray emission originating from these channels, namely the final-state radiation (FSR) single-photon emission, and the immediate hadronization and decay of the DM annihilation products into photons. In Refs.~\cite{Abazajian_GeV_excess,Daylan_GeV_excess}, the authors also added the bremsstrahlung contribution from electrons generated by the showering of the $b \bar{b}$ final state, but without taking electron diffusion into account. However, electrons produced in hadronization and decay processes do propagate in the Galaxy and eventually lose energy. The resulting population of electrons has an energy distribution slightly shifted towards the lower energy range but, depending on the energy propagation, is nevertheless expected to also emit photons in the GeV range through the bremsstrahlung process and inverse Compton scattering off the cosmic microwave background (CMB), UV and IR light, and starlight.

Here we show that the corresponding gamma-ray emission should not be neglected as it typically induces a signal in the energy range where the excess has been observed. The importance of the contribution from inverse Compton scattering was argued in Ref.~\cite{Fermi_IC} in the general context of setting constraints on DM annihilations from the diffuse gamma-ray emission from the Galaxy. However, here we show that these contributions from diffused electrons do not simply induce corrections to the gamma-ray spectrum, but in fact they drastically change the interpretation of the excess in terms of DM. More specifically, it turns out that one can fit the data very well with leptons in the final state, in particular with a pure leptonic final state. So far, these primary pure leptonic channels have been neglected in the literature because the associated prompt gamma-ray emission does not provide a good fit to the data \cite{GordonMacias}. However, our results show that the diffuse emission component originating from primary and secondary electrons should be considered very seriously, if the excess were indeed of DM origin.

In Sec.~\ref{prompt_vs_prop}, we recall the basics of the diffusion of electrons and remind the readers how these particles could contribute to the diffuse emission of gamma rays in our Galaxy. In Sec.~\ref{fits}, we fit the data and show how taking into account primary and secondary electrons can modify the interpretation of the GeV excess when the final state contains a large fraction of leptons. We provide a discussion of constraints from the AMS data and the morphology of the signal for leptophilic final states in Sec.~\ref{tests} and conclude in Sec.~\ref{conclusion}.

\section{Diffuse gamma-ray emission}
\label{prompt_vs_prop}

In this section, we describe the calculation of the additional contributions to the gamma-ray spectrum from electron diffusion.

\subsection{Prompt emission vs diffuse emission from primary and secondary electrons}

Since all the primary annihilation channels discussed in this paper contain charged particles, one expects a significant amount of prompt photon emission from both the FSR process and hadronization/decay of these primary particles (when hadronization can indeed take place).
 
Given that prompt emission is supposed to occur ``instantaneously," the corresponding gamma-ray signal offers a direct measurement of the DM energy distribution. As a result, any excess of gamma-ray photons that can be interpreted as mostly originating from prompt emission gives important information about the DM density profile and the decaying/annihilating nature of the DM.

However, when estimating the gamma-ray flux from these channels, one also needs to fold in the fact that they also eventually produce secondary electrons (in addition to primary electrons if the annihilation channel relies on a fraction of $e^+ e^-$). Since the latter propagate both spatially and in energy in the Galaxy, a consequence of these annihilation channels is the existence of a low-energy population of electrons whose origin is ultimately related to DM. This population is expected to produce diffuse gamma-ray emission due to its scattering off the interstellar radiation field (ISRF) and interactions with atomic nuclei in the interstellar medium. The associated spectrum may be a subcomponent of the total gamma-ray emission, but should definitely be taken into account since it could be an important element to understand the nature of DM. 

While this diffuse emission has no direct connection with the DM energy density distribution due to propagation, it can nevertheless be predicted from a given DM halo profile, provided we make some minimal assumptions about the efficiency of the inverse Compton (IC) scattering mechanism and bremsstrahlung.

\subsection{Propagation of electrons in the inner Galaxy}

To compute the gamma-ray spectrum from these DM-induced electrons, one first has to solve the diffusion-loss equation of cosmic rays, in order to compute the electron spectrum after propagation that enters into the expression of the gamma-ray flux.

\subsubsection{Transport equation}

Assuming a steady state, the diffusion-loss equation of cosmic rays reads \cite{Longair, Itilda, Synchrotron_wimps}
\begin{equation}
\label{transport equation}
K \ \nabla^{2}\psi \ + \ \dfrac{\partial}{\partial E}(b_{\mathrm{tot}}\ \psi) \ + \ q = 0,
\end{equation}
where $ \psi \equiv \psi ( \vec{x},E) $ is the electron spectrum (number density per unit energy) at location $ \vec{x} $ and energy $ E $. $ \nabla^{2} $ is the Laplacian operator, $q \equiv q(\vec{x},E) $ is the source term and $ b_{\mathrm{tot}} \equiv b_{\mathrm{tot}}(\vec{x},E) $ describes the total energy loss of the particle. The diffusion coefficient $ K $ models the transport through the small irregularities in the Galactic magnetic field. It is assumed to be independent of the position of the cosmic rays and is generally parametrized in the following way \cite{Itilda, Synchrotron_wimps, Synchrotron_DM_decay, Wechakama}: $ K(E) = K_{0} \left( E/E_{0} \right) ^{\delta} $, where $ E_{0} $ is an energy normalization taken to be $ 1 \ \rm GeV $. The diffusion model is defined by $ K_{0} $, $ \delta $, and the half-thickness $ L $ of the diffusion zone. Cosmic rays in the Milky Way Galaxy are indeed confined by the Galactic magnetic field to a diffusion zone modeled by a cylinder of radius $ R_{\mathrm{gal}} = 20 \ \rm kpc $ and half-thickness $ L $ with respect to the Galactic plane. The three sets of parameters we consider are given by
\begin{align}
\label{propa_sets}
\mathrm{MIN}:\ L & = 1\ \mathrm{kpc},K_{0} = 0.0016\ \rm kpc^{2}\ Myr^{-1},\delta = 0.85, \nonumber \\ 
\mathrm{MED}:\ L & = 4 \ \mathrm{kpc},K_{0} = 0.0112 \ \rm kpc^{2} \ Myr^{-1},\delta = 0.7, \nonumber \\
\mathrm{MAX}:\ L & = 15 \ \mathrm{kpc},K_{0} = 0.0765 \ \rm kpc^{2} \ Myr^{-1},\delta = 0.46,
\end{align}
with the medium (MED) set providing the best fit to the cosmic-ray measurements of the boron-to-carbon (B/C) ratio at the Earth's position \cite{Synchrotron_wimps}. The minimum (MIN) and maximum (MAX) sets allow one to quantify the uncertainties on the diffusion models compatible with observational data. 

Assuming that secondary and primary electrons only originate from DM annihilations, the source term reads
\begin{equation}
\label{source term}
q( \vec{x},E) = \dfrac{1}{\eta} \left\langle \sigma v \right\rangle \left( \dfrac{\rho ( \vec{x})}{m_{\mathrm{DM}}} \right) ^{2} \dfrac{\mathrm{d}n}{\mathrm{d}E}(E),
\end{equation}
where $ \left\langle \sigma v \right\rangle $ is the thermally averaged cross section times the relative velocity of the DM particles, $ \rho(\vec{x})$ is the DM density at position $\vec{x}$, $ m_{\mathrm{DM}} $ is the mass of the DM particles, and the numerical factor $ \eta $ accounts for the DM nature (Dirac vs Majorana or self-conjugate vs complex, i.e., $\eta = 4$ vs $\eta = 2$). We take $\eta = 2$ throughout this work. The term $ \mathrm{d}n/\mathrm{d}E $ represents the injection energy spectrum of electrons originating from the different channels of DM annihilations. 

In this paper, we take the same DM halo profile as in Ref.~\cite{GordonMacias}, namely the generalized Navarro-Frenk-White (NFW) profile:
\begin{equation}
\rho(r) = \rho_{\odot} \left( \dfrac{r}{r_{\odot}} \right)^{-\gamma}\left( \dfrac{1+\left( \dfrac{r}{r_{\mathrm{s}}} \right)^{\alpha}}{1+\left( \dfrac{r_{\odot}}{r_{\mathrm{s}}} \right)^{\alpha}} \right) ^{-(\beta-\gamma)/\alpha},
\end{equation}
with $\alpha=1$, $\beta=3$, $\gamma = 1.2$, $\rho_{\odot} = 0.36\ \rm GeV\ cm^{-3}$ and $r_{\mathrm{s}} = 23.1\ \rm kpc$. We also put a cutoff in the profile at $4.2 \times 10^{-7}\ \rm pc$, namely the Schwarzschild radius of the supermassive black hole at the GC. This value is about the same as that determined by the balance between accretion of DM particles onto the black hole and annihilations. We checked the consistency of this cutoff with the literature by reproducing the results of Refs.\cite{GordonMacias,Hooper} for the prompt emission. In practice a cutoff at a slightly larger scale should not make any difference in the results since the angular resolution of Fermi-LAT is not that good.

For the injection spectra, we make use of the values of $ \mathrm{d}n/\mathrm{d}E $ computed and tabulated for various DM masses in Ref.~\cite{Cirelli_cookbook} using the PYTHIA event generator \cite{Pythia}. 

To estimate the importance of electron diffusion, one needs to specify the different losses. At energies $\lesssim 10\ \rm GeV$, the main loss terms come from IC scattering on the different components of the ISRF (CMB, starlight, and IR and UV light), synchrotron radiation and bremsstrahlung emission. The synchrotron energy-loss term reads \cite{Longair}
\begin{equation}
b_{\mathrm{syn}} = \dfrac{4}{3} \sigma_{\mathrm{T}} c \dfrac{B^{2}}{2 \mu_{0}} \gamma_{\mathrm{L}} ^{2},
\end{equation}
where $ \sigma_{\mathrm{T}} $ is the Thomson cross section, $ B $ is the intensity of the magnetic field, $c$ is the speed of light, $\gamma_{\mathrm{L}}$ is the Lorentz factor of the electrons, and $ \mu_{0} $ is the vacuum permeability. 

The bremsstrahlung loss term depends on the species that compose the interstellar gas and on whether the matter is ionized or neutral. In this work, we consider for simplicity neutral hydrogen so the corresponding bremsstrahlung loss term in this strong-shielding limit reads \cite{Marco_brem,blumenthal}
\begin{equation}
b_{\mathrm{brems}} = \alpha_{\mathrm{em}} \dfrac{3 \sigma_{\mathrm{T}}}{8 \pi} n_{\mathrm{gas}} \left( \dfrac{4}{3} \phi^{\mathrm{H}}_{1,\mathrm{ss}} - \dfrac{1}{3} \phi^{\mathrm{H}}_{2,\mathrm{ss}} \right) E,
\end{equation}
where $\alpha_{\mathrm{em}}$ is the fine-structure constant, $E$ is the energy, $\phi^{\mathrm{H}}_{1,\mathrm{ss}} = 45.79$, and $\phi^{\mathrm{H}}_{2,\mathrm{ss}} = 44.46$.
The bremsstrahlung loss term therefore depends on the number density $n_{\mathrm{gas}}$ in the region of injection of the electrons. The authors of Ref.~\cite{Marco_brem} considered two models for the gas in the Galaxy with the density reaching values of ${\cal{O}}(1)$ or ${\cal{O}}(100)\ \rm cm^{-3}$. They used GALPROP maps \cite{galprop} that led to ${\cal{O}}(1)\ \rm cm^{-3}$ densities, and the density of ${\cal{O}}(100)\ \rm cm^{-3}$ corresponds to a toy model that relies on a modification of the GALPROP maps, by crudely taking into account the clumpiness of the gas distribution. In this work, we use a conservative approach and we only consider values of ${\cal{O}}(1)\ \rm cm^{-3}$ for $n_{\mathrm{gas}}$. A higher gas number density would increase the bremsstrahlung losses and thus the confinement of the electrons. Consequently this would increase the bremsstrahlung emission and reduce the IC contribution.

As for IC losses, we reproduce the calculation of Refs.~\cite{Timur_ISRF,myspike} which consists in fitting the ISRF spectrum with several greybody spectra corresponding to the different components (CMB, stars, IR, UV). Given that we are interested in a small region around the GC, we use the ISRF spectrum in the inner Galaxy. We thus consider homogeneous losses, but this should be a valid assumption since we focus on a small region around the GC. The corresponding losses $b_{\mathrm{IC}}$ are then computed for the different components in the different energy regimes as presented in Ref.~\cite{Timur_ISRF}. Finally, the total energy-loss term is given by $b_{\mathrm{tot}} = b_{\mathrm{syn}} + b_{\mathrm{brems}} + b_{\mathrm{IC}}$.

\subsubsection{Electron spectrum after diffusion}

We solve the transport equation by using the semianalytical method presented in Ref.~\cite{Itilda}. In this approach, the spectrum $ \psi $ of the cosmic-ray particle after propagation is given by the expression:
\begin{equation}
\psi (\vec{x},E) = \dfrac{\kappa}{b_{\mathrm{tot}}(E)} \int_{E}^{\infty} \! \tilde{I}_{\vec{x}}(\lambda_{\mathrm {D}}(E,E_{S})) \dfrac{\mathrm{d}n}{\mathrm{d}E}(E_{S}) \, \mathrm{d}E_{S}.
\end{equation}
In this expression $ \tilde{I}_{\vec{x}}(\lambda_{\mathrm {D}}(E,E_{S})) $ is the halo function, and $ \kappa = (1/2) \left\langle \sigma v \right\rangle (\rho_{\odot}/m_{\mathrm{DM}})^{2} $ is defined by writing the source term as $ q = \kappa (\rho/\rho_{\odot})^{2} \mathrm{d}n/\mathrm{d}E $. The halo function encodes all the information on diffusion through the diffusion length $ \lambda_{\mathrm {D}} $. The latter represents the distance traveled by a particle produced at energy $ E_{S} $ and losing energy during propagation, down to an energy $ E $. It is given by (see, e.g., Ref.~\cite{Synchrotron_wimps}):
\begin{equation}
\lambda_{\mathrm {D}}^{2}(E,E_{S}) = 4 \int_{E}^{E_{S}} \! \dfrac{K(E')}{b_{\mathrm{tot}}(E')} \, \mathrm{d}E'.
\end{equation}

To compute the halo function, we use the method relying on Green's functions detailed in Refs.~\cite{Green,myspike}. In this approach, the halo function is given by the convolution of the propagator $ G $ of the transport equation with the DM density squared, over the diffusion zone (DZ) \cite{Itilda},
\begin{equation}
\label{Itilde_convolution}
\tilde{I}_{\vec{x}}(\lambda_{\mathrm {D}}(E,E_{S})) = \int_{\mathrm{DZ}} \! \mathrm{d}\vec{x}_{\mathrm{S}} \, G(\vec{x},E \leftarrow \vec{x}_{\mathrm{S}},E_{\mathrm{S}}) \left( \dfrac{\rho(\vec{x}_{\mathrm{S}})}{\rho_{\odot}} \right) ^{2}.
\end{equation}
However, for small values of $ \lambda_{\mathrm {D}} $ relative to the distance from the GC, the propagator becomes very sharply peaked. Moreover, the DM profile is also very sharply peaked close to the GC. Consequently, if the integrand is not correctly sampled, the halo function is underestimated at the GC. To solve this issue, we use logarithmic steps to account for the cuspiness of the profile, but the sharpness of the propagator nevertheless requires a more complex treatment which is given in  Ref.~\cite{myspike}.

\section{Fitting the GeV excess}
\label{fits}

Using this dedicated treatment of diffusion on very small scales, we can now estimate the relative importance of the diffuse gamma-ray emission generated through the propagation of secondary (and primary) electrons with respect to the prompt emission, and how this additional contribution affects the fit to the Fermi-LAT excess. We consider three specific scenarios in which DM particles annihilate either into 100$\%$ leptons, a mixture of leptons and $b\bar{b}$ or 100$\%$ $b \bar{b}$. We recall the fact that ``leptons" refers to a mixture of the $e^{+}e^{-}$, $\mu^{+}\mu^{-}$, $\tau^{+}\tau^{-}$ channels, with 1/3 of the annihilations into each of these channels.

\subsection{Prompt, IC and bremsstrahlung contributions}

To compare the importance of the different components, we  use a $7^{\circ} \times 7^{\circ}$ region corresponding to the signal found in Ref.~\cite{GordonMacias}.

\subsubsection{Prompt emission}

The flux of prompt gamma rays (energy per unit time per unit surface area per unit solid angle) is given by the integral over the line of sight coordinate $s$ of the DM density squared (see, e.g., Ref.~\cite{Bringmann}),
\begin{equation}
E_{\gamma}^{2}\dfrac{\mathrm{d}n}{\mathrm{d}E_{\gamma}\mathrm{d}\Omega} = \dfrac{E_{\gamma}^{2}}{4\pi} \dfrac{1}{2} \left( \dfrac{\rho_{\odot}}{m_{\mathrm{DM}}}\right) ^{2} \left\langle \sigma v \right\rangle \dfrac{\mathrm{d}N}{\mathrm{d}E_{\gamma}} \int_{\mathrm{l.o.s.}} \! \left( \dfrac{\rho(\vec{x})}{\rho_{\odot}}\right) ^{2} \, \mathrm{d}s
\end{equation}

The  flux from the squared $7^{\circ} \times 7^{\circ}$ region is then given by
\begin{equation}
E_{\gamma}^{2}\dfrac{\mathrm{d}n}{\mathrm{d}E_{\gamma}} = 4 \int_{0}^{\theta_{\mathrm{fov}}} \int_{0}^{\theta_{\mathrm{fov}}} \! E_{\gamma}^{2}\dfrac{\mathrm{d}n}{\mathrm{d}E_{\gamma}\mathrm{d}\Omega} \cos b \, \mathrm{d}b \ \mathrm{d}l,
\end{equation}
where $l$ and $b$ are, respectively, the longitude and the latitude, and $\theta_{\mathrm{fov}} = 3.5^{\circ}$ defines the field of view. This corresponds to the flux expected for a given annihilation channel. To get the total flux, we then sum and  weight the different channels (leptons, leptons+$b$ quarks, $b\bar{b}$).

\subsubsection{IC and bremsstrahlung emissions}

In contrast, computing the flux of gamma rays emitted by electrons requires taking propagation into account. This can be expressed as (see, e.g., Ref.~\cite{Cirelli_cookbook})
\begin{equation}
\label{flux}
E_{\gamma}^{2}\dfrac{\mathrm{d}n}{\mathrm{d}E_{\gamma}\mathrm{d}\Omega} = \dfrac{E_{\gamma}}{4\pi} \int_{\mathrm{l.o.s.}} \! j(E_{\gamma},s,l,b) \, \mathrm{d}s,
\end{equation}
where $j(E_{\gamma},s,l,b) \equiv j(E_{\gamma},\vec{x})$ is the photon emissivity (power per unit volume per unit energy) obtained after propagation of the electrons and after taking into account the photon emission due to their interactions with the ISRF and atomic nuclei in the interstellar medium. The emissivity is therefore given by (see Refs.~\cite{blumenthal,Marco_brem})
\begin{equation}
\label{emissivity}
j(E_{\gamma},\vec{x}) = N_{\mathrm{e}} \int_{E_{\mathrm{e}}^{\mathrm{min}}}^{m_{\mathrm{DM}}} \! P(E_{\gamma},E_{\mathrm{e}},\vec{x}) \psi(E_{\mathrm{e}},\vec{x}) \, \mathrm{d}E_{\mathrm{e}},
\end{equation}
where $\psi$ is the electron spectrum after propagation, $P = P_{\mathrm{IC}} + P_{\mathrm{brems}}$ is the emission spectrum, $N_{\mathrm{e}} = 2$ takes into account the fact that both electrons and positrons radiate, and $E_{\mathrm{e}}^{\mathrm{min}}$ is the minimum electron energy from kinematics.

\begin{figure}[t]
\centering
\includegraphics[scale=0.44]{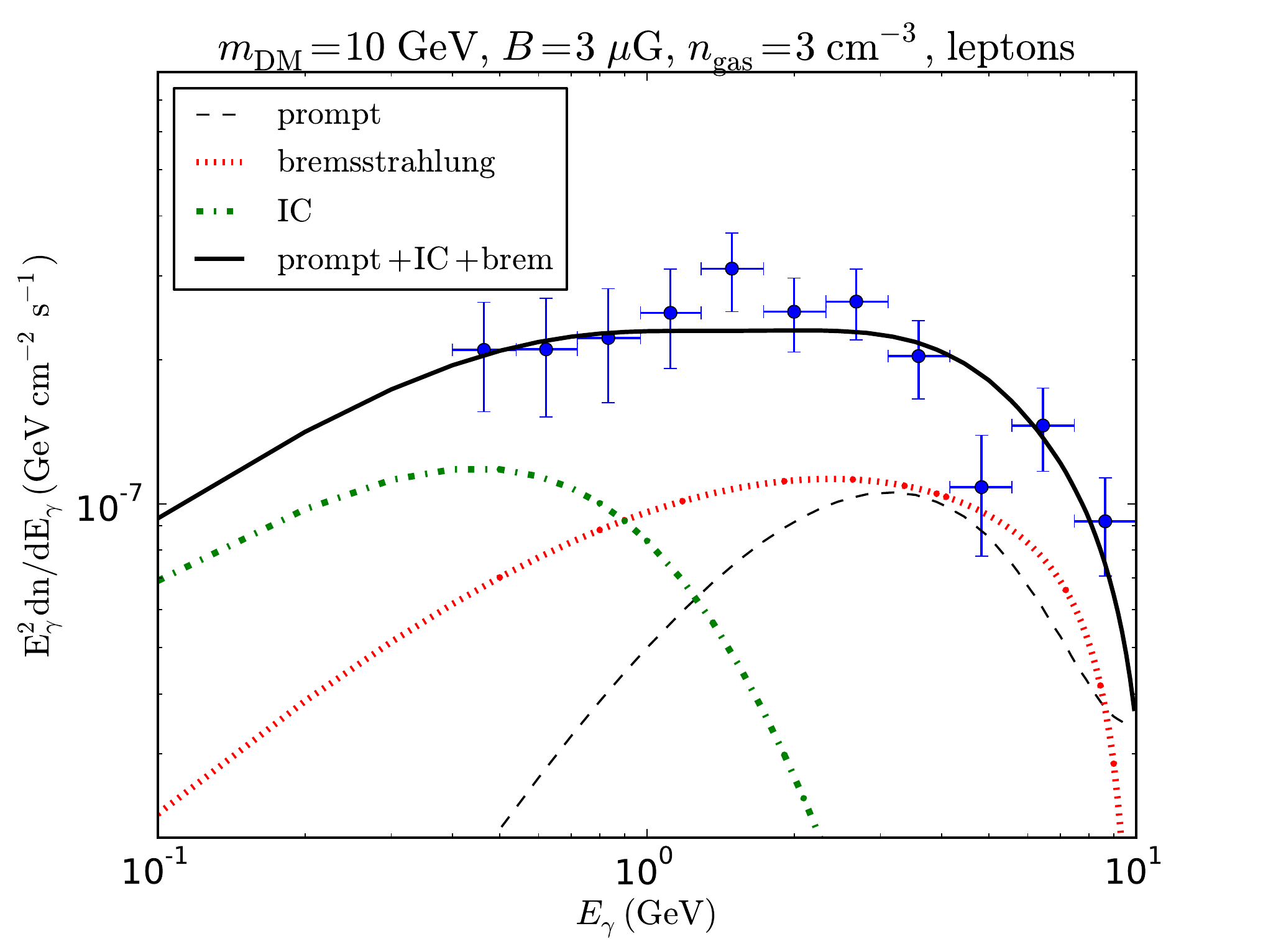}
\caption{\label{leptons_3contributions}Spectrum of the residual extended emission in the $7^{\circ} \times 7^{\circ}$ region around the GC. The blue points are the residuals in the Fermi-LAT data extracted by the authors of Ref.~\cite{GordonMacias}. The prompt contribution (black dashed), IC (green dashed-dotted) and bremsstrahlung (red dotted) emissions from $10\ \rm GeV$ DM annihilating only into leptons democratically add up to give a very good fit to the data, as shown by the black solid line. This figure is obtained for a best-fit cross section of $\left\langle \sigma v \right\rangle = 0.86 \times 10^{-26}\ \rm cm^{3}\ s^{-1}$. }
\end{figure}

\begin{figure*}[t]
\centering
\includegraphics[scale=0.43]{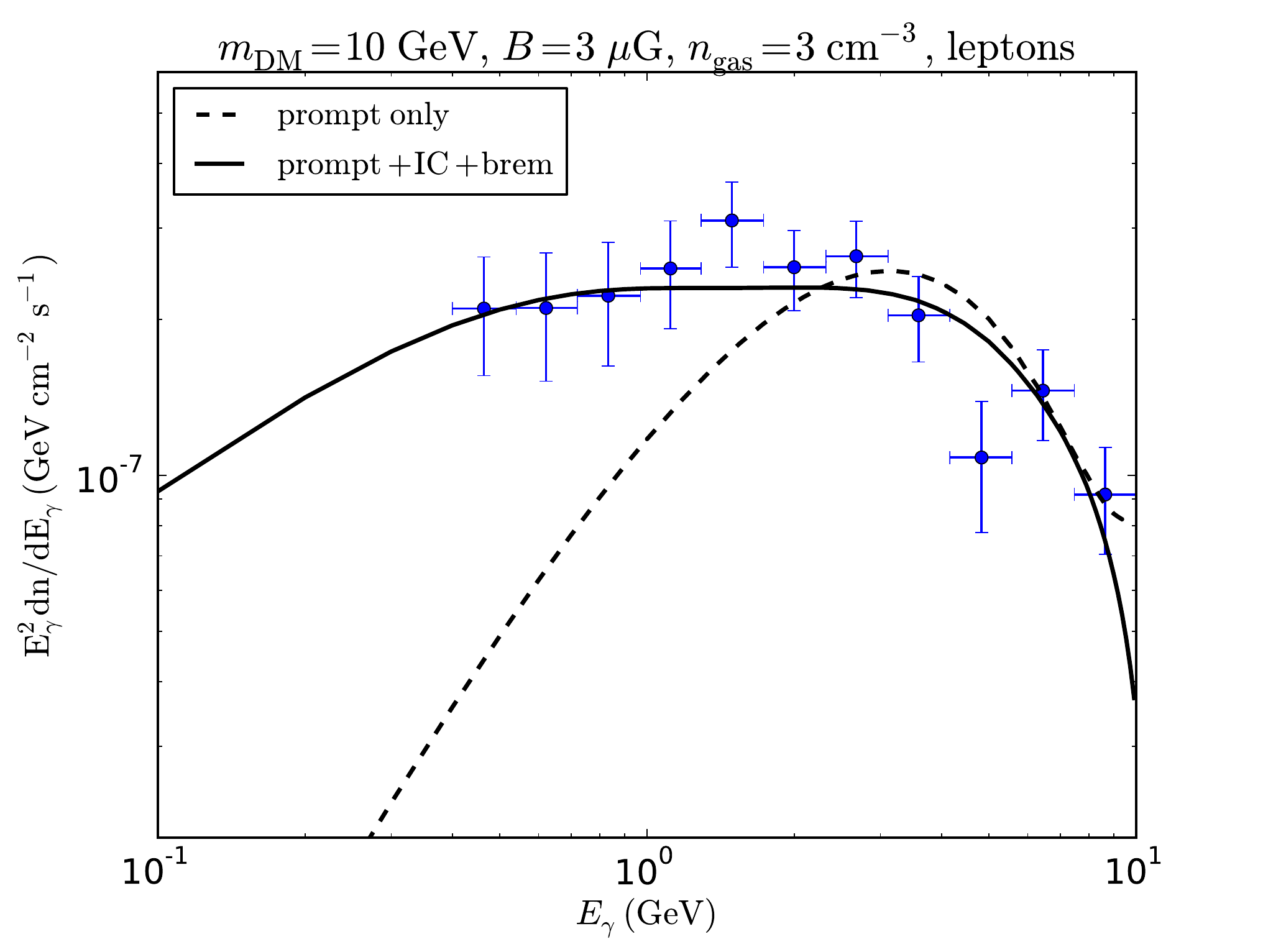}
\includegraphics[scale=0.43]{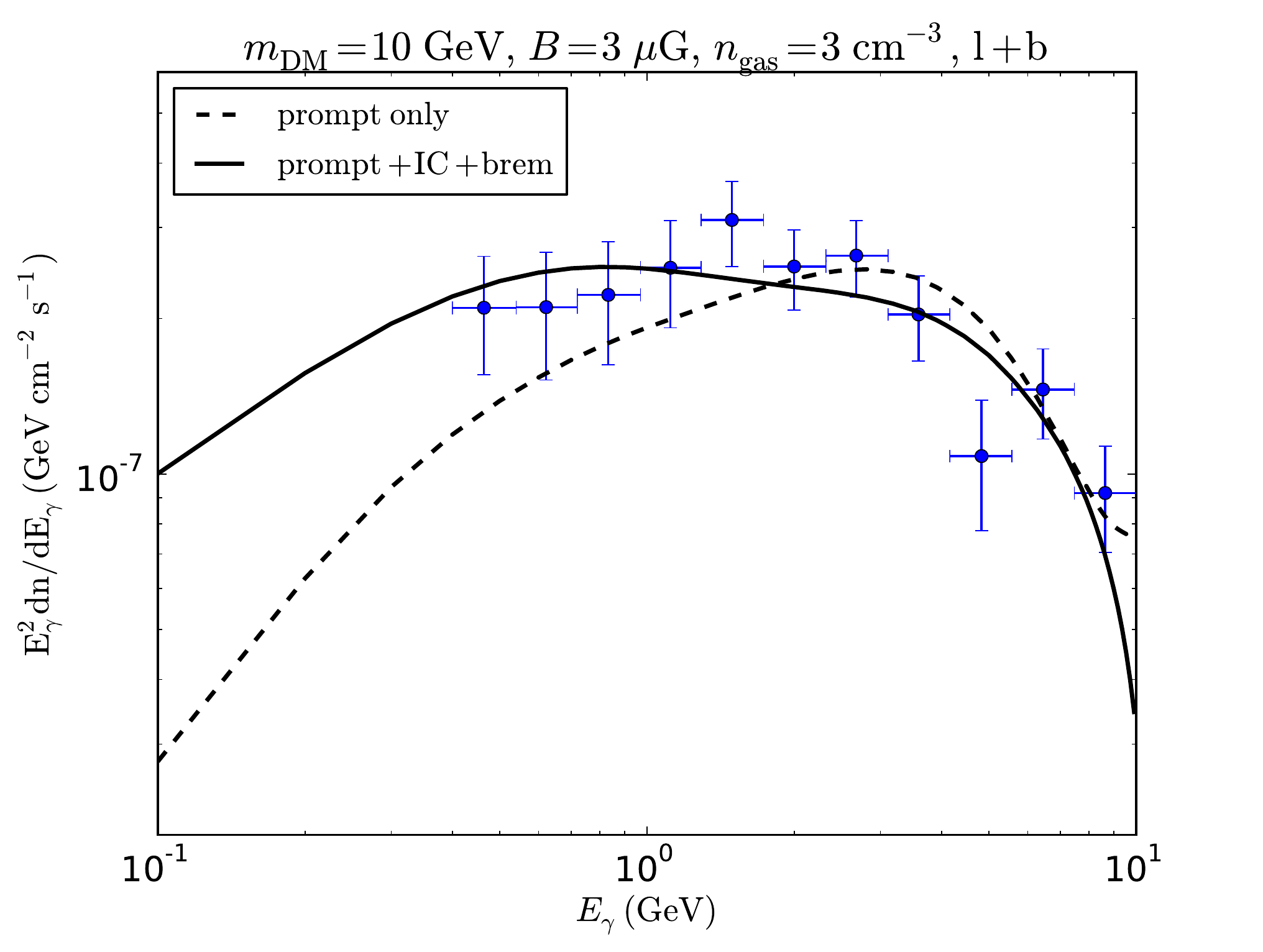}
\caption{\label{leptons_best_fits}Best fits to the Fermi residual with the gamma-ray spectrum from annihilations of $10\ \rm GeV$ DM particles into leptons democratically (left panel) or 90\% leptons and 10\% $b$ quarks (right panel). In both panels we show the best fit with only prompt gamma-ray emission and with the three contributions from prompt, IC and bremsstrahlung emissions. The corresponding best-fit cross sections for the total spectrum are about the same for leptons and leptons+$b$ quarks, i.e., respectively, $\left\langle \sigma v \right\rangle = 0.86 \times 10^{-26}\ \rm cm^{3}\ s^{-1}$ and $\left\langle \sigma v \right\rangle = 0.89 \times 10^{-26}\ \rm cm^{3}\ s^{-1}$. When fitting the data with only prompt gamma rays, the cross section is $2.02 \times 10^{-26}\ \rm cm^{3}\ s^{-1}$ for leptons and $2.11 \times 10^{-26}\ \rm cm^{3}\ s^{-1}$ for leptons+$b$ quarks. For leptons only, including the gamma-ray emission from diffused electrons significantly improves the fit.}
\end{figure*}

For IC emission, the emission spectrum reads (see e.g., Refs.~\cite{blumenthal,Marco_brem})
\begin{align}
\label{P_IC}
P_{\mathrm{IC}}(E_{\gamma},E_{\mathrm{e}},\vec{x}) = \dfrac{3 \sigma_{\mathrm{T}} c}{4 \gamma_{\mathrm{L}}^{2}} \int_{1/4\gamma_{\mathrm{L}}^{2}}^{1} \! \mathrm{d}q \left( E_{\gamma} - E_{\gamma}^{0}(q) \right) \dfrac{n(E_{\gamma}^{0}(q),\vec{x})}{q} \nonumber \\ 
\times \left( 2q\ln q + q + 1 -2q^{2} +\dfrac{1}{2} \dfrac{\epsilon^{2}}{1-\epsilon} (1-q)\right),
\end{align}
where $\epsilon = E_{\gamma}/E_{\mathrm{e}}$ and the initial energy of the photon of the ISRF is related to $q$ via:
\begin{equation}
E_{\gamma}^{0}(q) = \dfrac{E_{\gamma}}{4 q \gamma_{\mathrm{L}}^{2} (1-\epsilon)}.
\end{equation}
In Eq.~\eqref{P_IC}, $n$ is the sum of the number densities per unit energy for the different components of the photon bath. We assume a constant value for $n$, corresponding to the value at the GC. Note that the lower bound of the integral in Eq.~\eqref{emissivity} is equal to a minimum energy that is close to the energy of the emitted photon: $E_{\mathrm{e}}^{\mathrm{min}} = \left( E_{\gamma} + \left(E_{\gamma}^{2}+m_{\mathrm{e}}^{2}\right) ^{1/2}\right) /2$. For the gamma-ray energies of interest here (typically $E_{\gamma} > 0.1\ \rm GeV$), $E_{\mathrm{e}}^{\mathrm{min}}$ is very close to $E_{\gamma}$.

For bremsstrahlung emission, the spectrum is given by \cite{blumenthal,Marco_brem} (and multiplied by $E_{\gamma}$ to get a power per unit energy)
\begin{equation}
P_{\mathrm{brems}}(E_{\gamma},E_{\mathrm{e}},\vec{x}) = c n_{\mathrm{gas}} E_{\gamma} \dfrac{\mathrm{d}\sigma}{\mathrm{d}E_{\gamma}}(E_{\gamma},E_{\mathrm{e}}),
\end{equation}
with the differential cross section given by
\begin{equation}
\dfrac{\mathrm{d}\sigma}{\mathrm{d}E_{\gamma}} = \dfrac{3 \alpha_{\mathrm{em}} \sigma_{\mathrm{T}}}{8 \pi E_{\gamma}} \left[ \left( 1 + \left( 1 - \dfrac{E_{\gamma}}{E_{\mathrm{e}}} \right) ^{2}\right) \phi_{1} - \dfrac{2}{3} \left( 1 - \dfrac{E_{\gamma}}{E_{\mathrm{e}}} \right) \phi_{2}\right],
\end{equation}
with $\phi_{1} = \phi^{\mathrm{H}}_{1,\mathrm{ss}}$ and $\phi_{2} = \phi^{\mathrm{H}}_{2,\mathrm{ss}}$ when considering only atomic neutral hydrogen. In this case the lower bound for the integral over the electron energy is simply $E_{\gamma}$.

\subsection{Fits to the data}

We can now fit the Fermi-LAT excess. To make our point, we choose fixed values of the parameters describing the interstellar medium (in particular the magnetic field and gas density) and allow the annihilation cross section to vary. For simplicity, we assume the same value of the annihilation cross section for all the final states considered in this paper.

In principle one should scan over all possible free parameters (including in fact the magnetic field and gas density) but since we are only interested in showing that 10 GeV DM annihilating into a large fraction of leptons fits the data very well if one accounts for the diffusion and gamma-ray emission of the electrons, we keep a simplified setup with $B = 3\ \mu \rm G$ and $n_{\mathrm{gas}} = 3\ \rm cm^{-3}$. 

The data are taken from Ref.~\cite{GordonMacias} and correspond to a $7^{\circ} \times 7^{\circ}$ region. Different data sets were given in Refs.~\cite{Hooper,Abazajian_GeV_excess,Daylan_GeV_excess}, depending on the assumption on the background sources that are being subtracted from the data, but the implication of these different sets on the best-fit parameters is beyond the scope of the present paper.

\begin{figure}[ht]
\centering
\includegraphics[scale=0.44]{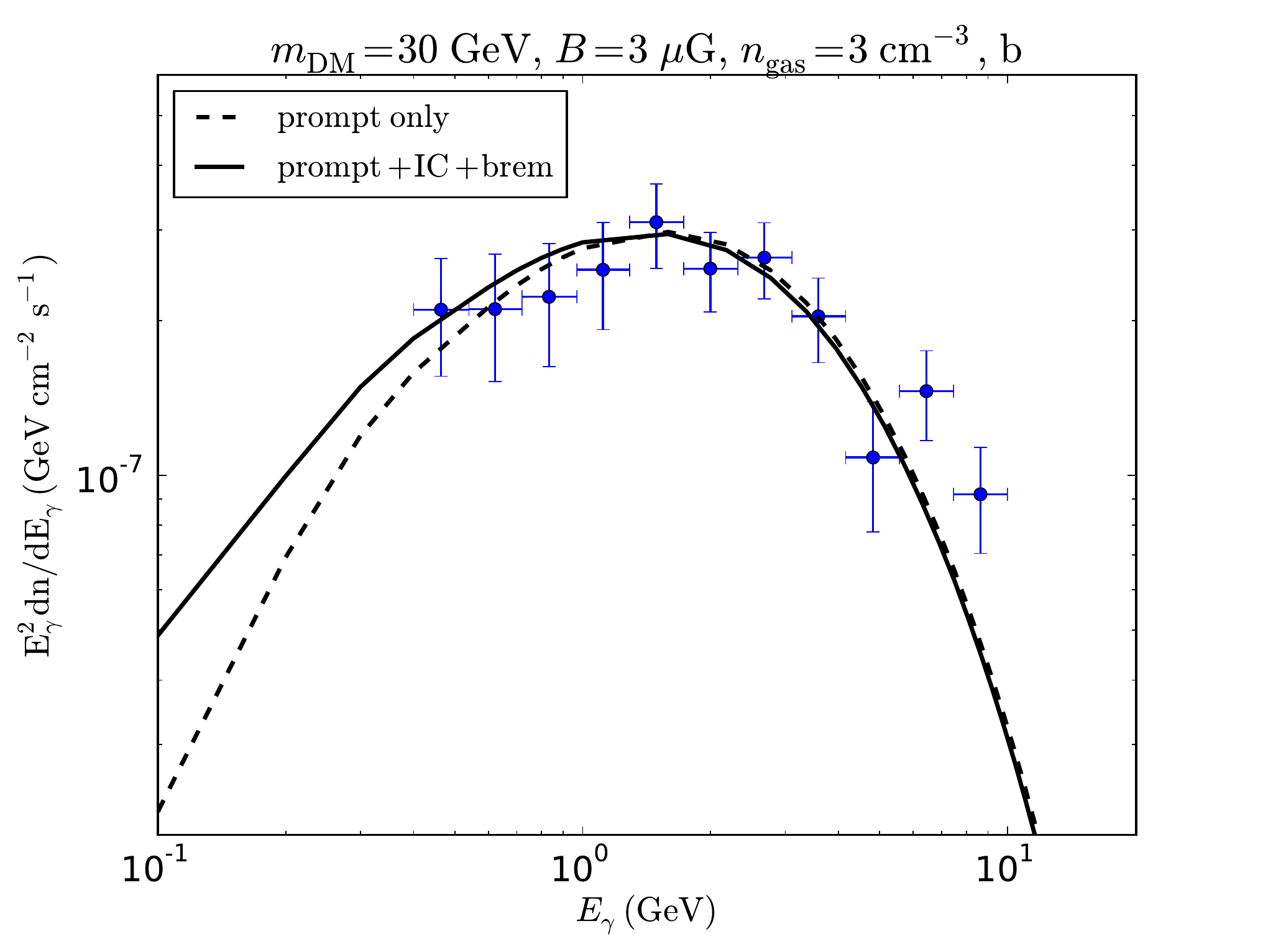} 
\caption{\label{b_best_fit}Best fits to the Fermi residual with the gamma-ray spectrum from annihilations of $30\ \rm GeV$ DM particles into 100\% $b\bar{b}$. Including the contributions from IC and bremsstrahlung emissions does not significantly affect the spectrum, except at low energies. The best-fit cross section in both cases is of the order of $2 \times 10^{-26}\ \rm cm^{3}\ s^{-1}$.}
\end{figure}

\begin{figure}[t]
\centering
\includegraphics[scale=0.44]{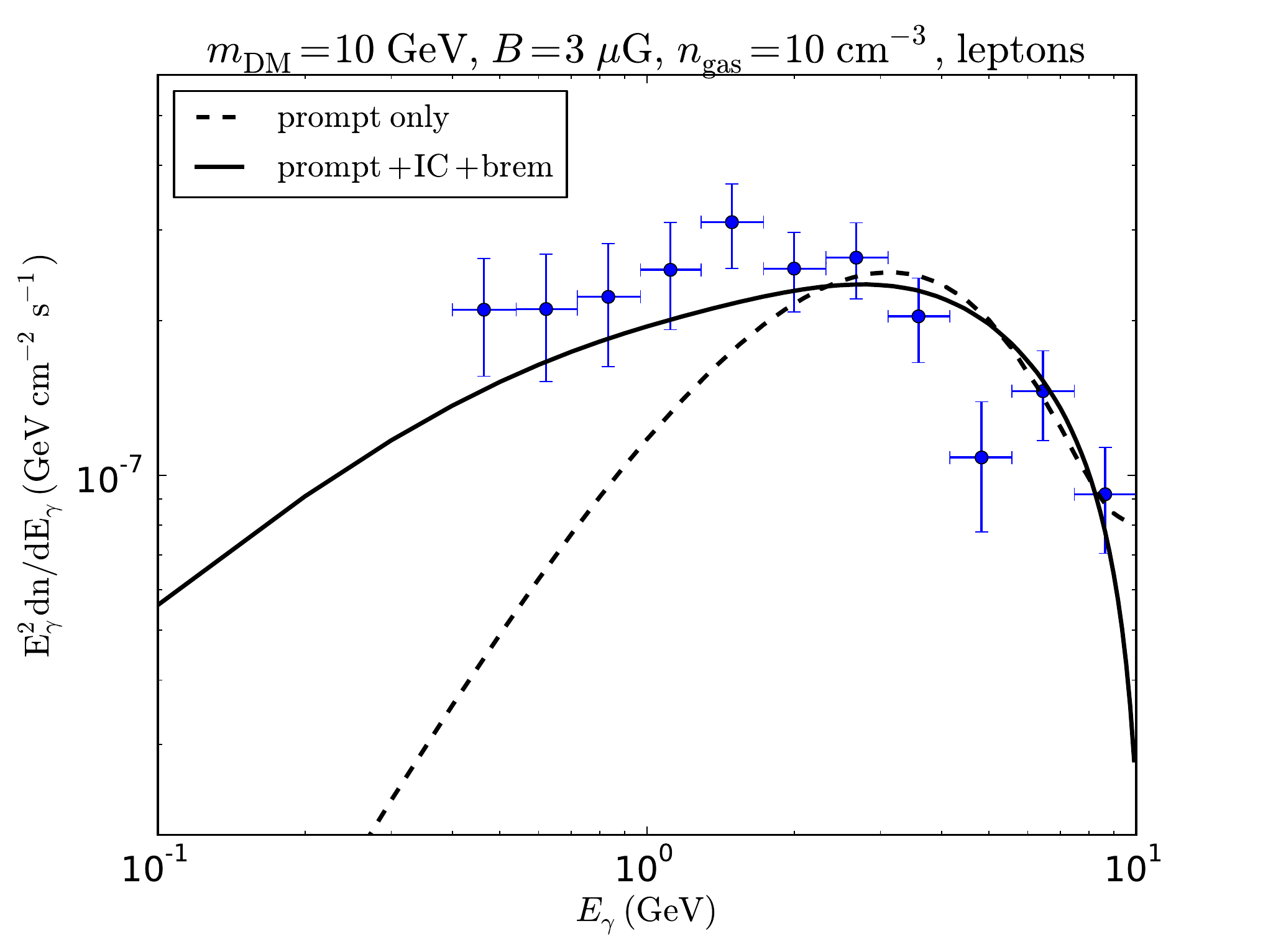} 
\caption{\label{n_gas_10}Best fits to the Fermi residual with the gamma-ray spectrum from annihilations of $10\ \rm GeV$ DM particles into leptons, with a gas density of $10\ \rm cm^{-3}$.}
\end{figure}

In Fig.~\ref{leptons_3contributions}, we compare the contributions from the prompt emission with that from IC and bremsstrahlung, for a pure leptonic channel. As one can readily see, IC emission is particularly important at low energy (below $1\ \rm GeV$) while bremsstrahlung emission is important at intermediate energies (1--10 GeV). Also, as pointed out already in previous work \cite{GordonMacias}, the prompt emission alone does not fit the data for the leptonic channel. However, our work shows that the sum of the three components (prompt, IC and bremsstrahlung) actually provides an excellent fit. 

In Fig.~\ref{leptons_best_fits}, we compare the best fits obtained with only prompt emission and prompt$+$IC$+$bremsstrahlung emissions, for a pure leptonic final state (left panel) and a scenario containing 90\% leptons and 10\% $b$ quarks (right panel). The importance of the IC and bremsstrahlung contributions is less crucial when DM can annihilate into $b \bar{b}$. Nevertheless, these IC and bremsstrahlung components enable one to significantly improve the quality of the fit. 

To make a more quantitative statement, we define the goodness of fit by the criterion $\chi^2 < 29.6$, which gives  a p-value greater than $10^{-3}$ \cite{pdg}, corresponding to 11 data energy bins and one free parameter, $\left\langle \sigma v \right\rangle$. Note that in our analysis we combine in quadrature the statistical and systematic errors provided in Ref.~\cite{GordonMacias}. For prompt emission with only leptons, the best fit is obtained for $\left\langle \sigma v \right\rangle = 2.02 \times 10^{-26}\ \rm cm^{3}\ s^{-1}$, with $\chi^2 = 41.93$, which is a very bad fit. However, we obtain a $\chi^2$ of 10.21 for a cross section of $0.86 \times 10^{-26}\ \rm cm^{3}\ s^{-1}$ when we add up the IC and bremsstrahlung contributions. This demonstrates the importance of taking into account the gamma-ray emission from diffused electrons. Note that the error bars on the cross section at the $1\sigma$ level are of the order of $0.06 \times 10^{-26}\ \rm cm^{3}\ s^{-1}$.

For the channel with 90\% leptons + 10\% $b\bar{b}$, the difference is smaller than for leptons only, but the $\chi^2$ is nevertheless reduced from 16.46 (with a best-fit cross section of $2.11 \times 10^{-26}\ \rm cm^{3}\ s^{-1}$) down to 9.57 (with a best-fit cross section of $0.89 \times 10^{-26}\ \rm cm^{3}\ s^{-1}$) when including IC and bremsstrahlung emissions. Hence, in such a scenario, both spectra with or without the IC and bremsstrahlung contributions fit the data, but there is a clear preference for the total spectrum.

\begin{figure}[t]
\centering
\includegraphics[scale=0.44]{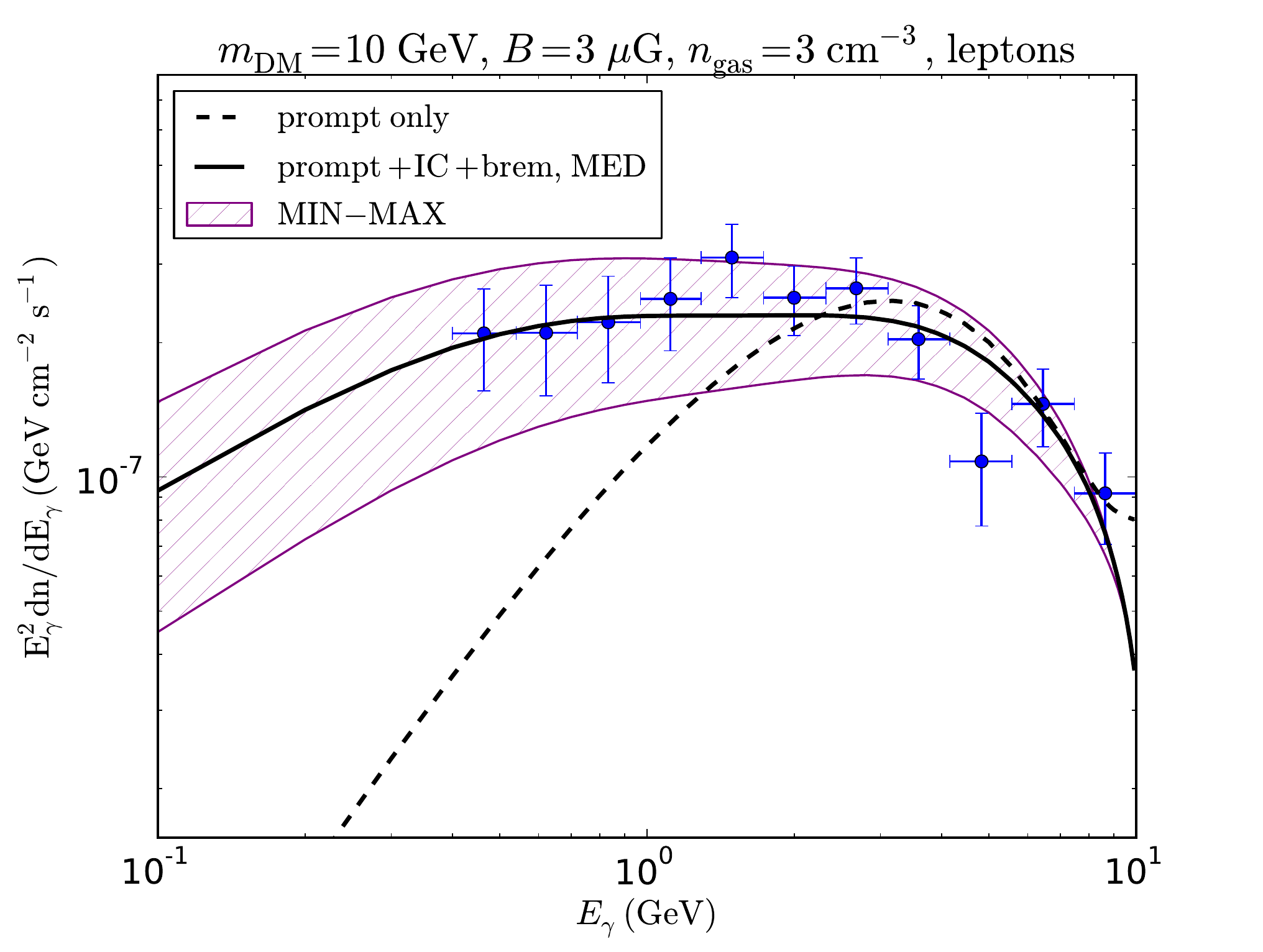} 
\caption{\label{min-max}Best fits to the Fermi residual with the gamma-ray spectrum from annihilations of $10\ \rm GeV$ DM particles into leptons. The purple hatched area represents the uncertainty on the best fit for the total spectrum including IC and bremsstrahlung due to the uncertainty on the diffusion model. The band is bracketed by the fluxes for the MIN and MAX sets, respectively, at the top and the bottom.}
\end{figure}

\begin{figure}[t]
\centering
\includegraphics[scale=0.44]{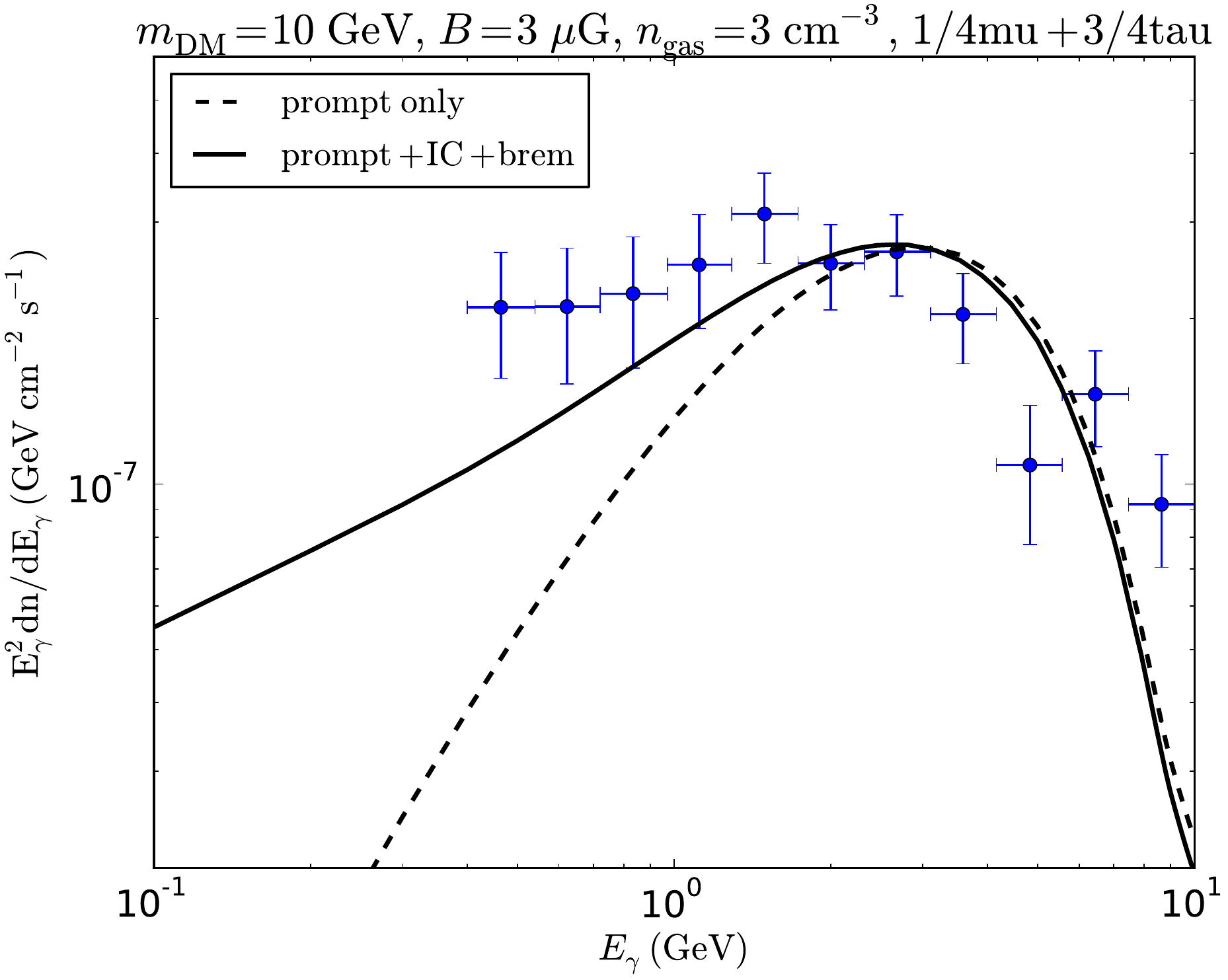} 
\caption{\label{mu_tau_best_fit}Best fits to the Fermi residual with the gamma-ray spectrum from annihilations of $10\ \rm GeV$ DM particles with branching ratios of 0.25 into muons and 0.75 into taus. The best-fit cross section is $\sim 1 \times 10^{-26}\ \rm cm^{3}\ s^{-1}$.}
\end{figure}

\begin{figure*}[t]
\centering
\includegraphics[scale=0.43]{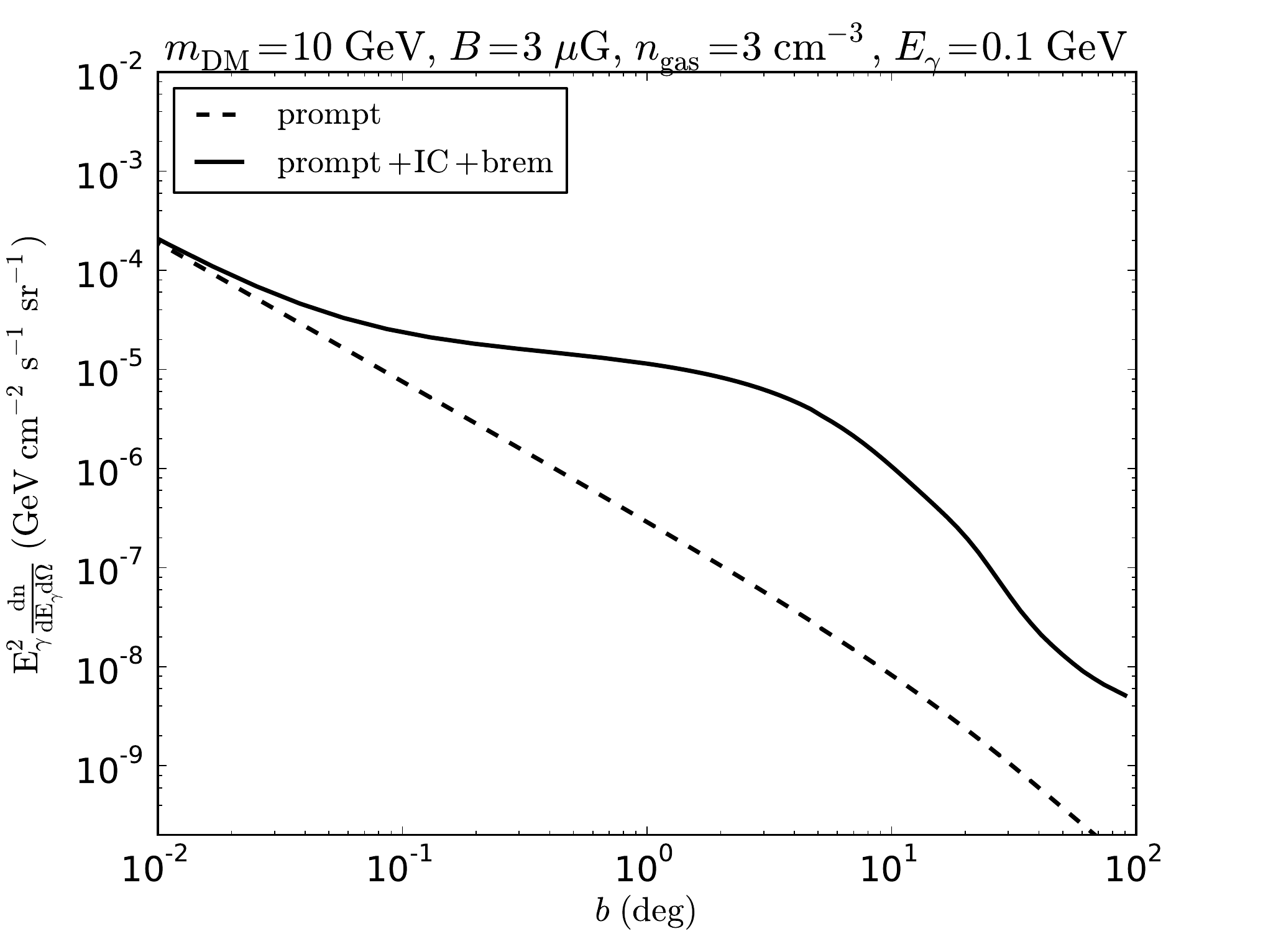}\includegraphics[scale=0.43]{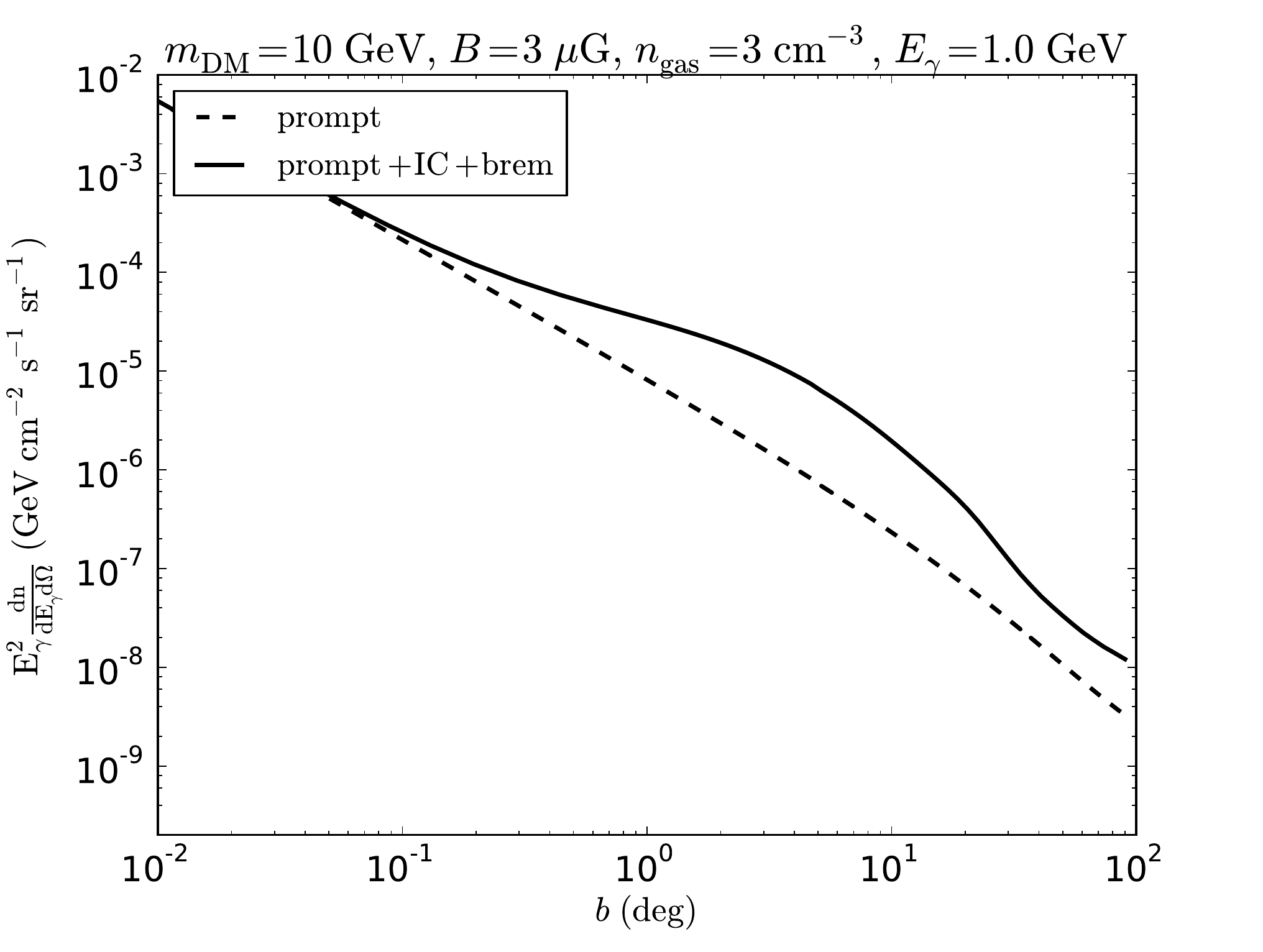} \includegraphics[scale=0.43]{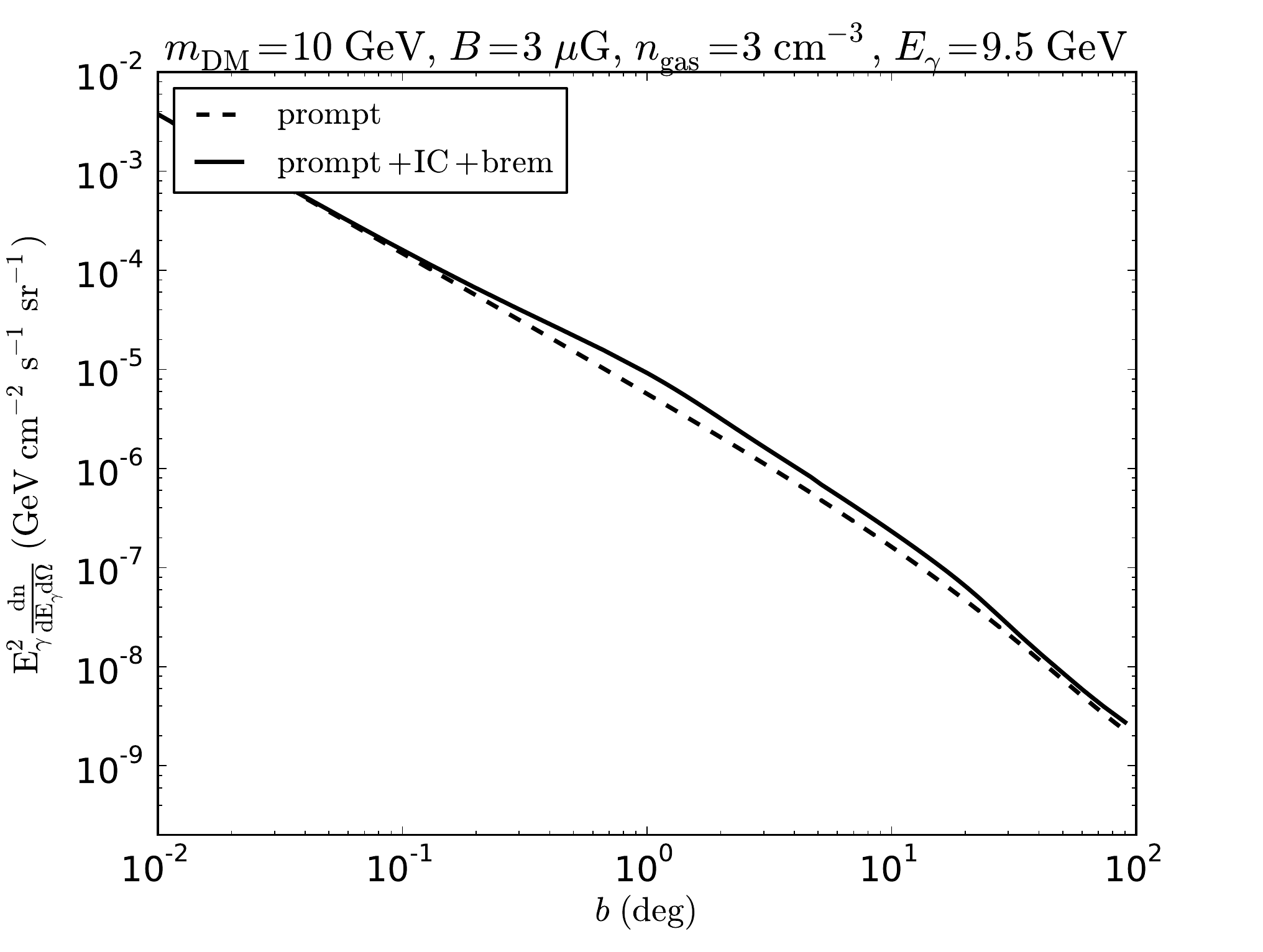} 
\caption{\label{flux_vs_b}Gamma-ray flux from DM annihilating exclusively into leptons, as a function of latitude $b$, for three values of the gamma-ray energy $E_{\gamma}$: 0.1, 1, and 9.5 GeV. 
The contributions from electron diffusion, via IC and bremsstrahlung, clearly lead to an excess with respect to the prompt emission below $10^{\circ}$ at low energy.}
\end{figure*}

Shown in Fig.~\ref{b_best_fit} are the best fits for the prompt spectrum and the total spectrum in the case of a 30 GeV DM particle annihilating into 100\% $b\bar{b}$. The corresponding best-fit values of the annihilation cross section are not very different: $\left\langle \sigma v \right\rangle = 2.2 \times 10^{-26}\ \rm cm^{3}\ s^{-1}$ for the prompt emission (with $\chi^2 = 11.24$), and $\left\langle \sigma v \right\rangle = 2.03 \times 10^{-26}\ \rm cm^{3}\ s^{-1}$ for the total emission 
(with $\chi^2 = 11.98$). In this case, the contributions from IC and bremsstrahlung are subdominant, except at low energy. This is due to the fact that the IC and bremsstrahlung emission spectra take large values for electron energies close to the DM mass ($E_{\mathrm{e}}$ must be much greater than  the observed energy $E_{\gamma}$). Electrons originating from $b\bar{b}$ tend to have an energy spectrum peaked at low energy, unlike those originating from leptonic annihilation channels that peak closer to the DM mass. Hence, looking at the gamma-ray spectrum at lower energies could be a good way to test whether the $b\bar{b}$ channel, which is usually claimed to be the preferred channel, indeed agrees with other data sets from the GC.

So far, we have shown that taking $B = 3\ \mu \rm G$ and $n_{\mathrm{gas}} = 3\ \rm cm^{-3}$ leads to a very good fit to the data with the total spectrum, particularly for the leptonic channel. However, the fits are fairly robust with respect to changes in these parameters. For instance, taking $B = 10\ \mu \rm G$---a value that may be more consistent with the value close to the GC---leads to a small global shift of the IC and bremsstrahlung contributions (due to greater losses). The resulting best fit is only slightly affected, with $\chi^2 = 10.35$ and $\left\langle \sigma v \right\rangle = 0.92 \times 10^{-26}\ \rm cm^{3}\ s^{-1}$ for the leptonic channel. When taking a greater value for $n_{\mathrm{gas}}$, namely $10\ \rm cm^{-3}$, the resulting spectrum is harder at low energy but still provides a very good fit to the data, with $\chi^2 = 16.6$ and $\left\langle \sigma v \right\rangle = 0.6 \times 10^{-26}\ \rm cm^{3}\ s^{-1}$, as shown in Fig.~\ref{n_gas_10}.

Finally, the diffusion model introduces an additional uncertainty, which is quantified by the MIN and MAX sets of propagation parameters and degenerated with the cross section (although changing the diffusion parameters mostly affects the low-energy end of the spectrum, since the prompt contribution remains fixed). This uncertainty is shown in Fig.~\ref{min-max}. The hatched area is bounded by the spectra for the MIN and MAX sets (respectively at the top and the bottom of the band) computed with the best-fit cross section obtained with the MED set. Hence the uncertainty on the diffusion model translates into an error on the best-fit value for the cross section. The corresponding values for the MIN and MAX sets are $\left\langle \sigma v \right\rangle_{\mathrm{MIN}} = 0.68 \times 10^{-26}\ \rm cm^{3}\ s^{-1}$ and $\left\langle \sigma v \right\rangle_{\mathrm{MAX}} = 1.18 \times 10^{-26}\ \rm cm^{3}\ s^{-1}$.

\section{Further tests}
\label{tests}

\subsection{Discussion of constraints from the AMS data}

We found two best fits in the leptonic case: one corresponding to the democratic scenario ($\chi^2=10.21$) and one without electrons and with branching ratios of about 2/3 into $\mu^{+}\mu^{-}$  and 1/3 into $\tau^{+}\tau^{-}$  ($\chi^2=14.22$). For the latter case the fit requires a cross section of $1.42 \times 10^{-26}\ \rm cm^{3}\ s^{-1}$.  The democratic scenario is however in tension with the limits on the annihilation cross section into $e^{+}e^{-}$ derived from the AMS data in Refs.~\citep{Hooper_AMS,Ibarra_AMS,Bringmann_constraints}, which essentially exclude annihilations into $e^{+}e^{-}$ with cross sections close to the thermal value. These limits also exclude branching ratios into $\mu^{+}\mu^{-}$ larger than 0.25~\cite{Bringmann_constraints}.  

The constraints from Ref~\cite{Bringmann_constraints} were obtained by searching for tiny deviations from a power-law background that empirically fits the AMS data. These limits would probably be less stringent if one relaxes the assumption of a smooth background. However, as shown in Fig.~\ref{mu_tau_best_fit}, when taking these constraints into account, we find that the effect of IC and bremsstrahlung becomes less significant than for a larger branching ratio into muons. The associated best-fit cross section for a branching ratio into muons of 0.25 is $\left\langle \sigma v \right\rangle \approx 1 \times 10^{-26}\ \rm cm^{3}\ s^{-1}$ and we find $\chi^2 = 27.3$, which corresponds to a marginally good fit.

\subsection{Morphology}

The morphology of the diffuse emission in the case of the democratic scenario depends on the observed energy. In Fig.~\ref{flux_vs_b}, we show the expected gamma-ray flux as a function of latitude (or similarly longitude) for three different energies (0.1 GeV, 1.0 GeV and 9.5 GeV). As one can see, secondary electrons can induce a significant excess of gamma-ray emission at low energies (below a few GeV) with respect to prompt emission. This contribution leads to a significant flux up to a few tens of degrees which is in agreement with Ref.~\cite{Daylan_GeV_excess}, where the authors found that the excess extends out to at least $12^{\circ}$.

Below $1\ \rm GeV$ (typically $0.1\ \rm GeV$), the diffusion contribution dominates over the contribution from prompt emission. Between 3 and $12^{\circ}$, we find that our model is well fitted by a power law with index 1.34, which is very close to the index of 1.4 that one obtains for prompt emission only, corresponding to a DM profile with a power-law index of 1.2. Therefore, in this energy range our model is consistent with the morphology of the prompt emission found in the literature (e.g., Ref.~\cite{Daylan_GeV_excess}). However, at such low energies (i.e., essentially 0.1 GeV), the diffusion contribution leads to a  profile for the flux between ${\cal{O}}(0.1)$ and ${\cal{O}}(1)^{\circ}$ that is shallower than the profile from prompt emission. At 1 GeV and for the same angular region, the tension is much weaker. But  in any case one should rather consider the results from the analysis of the Fermi signal that excludes the inner $1^{\circ}$ as this is more robust.

Investigating the morphology in the $ [ 0.1^{\circ},1^{\circ}]$ region, at energies below 0.1 GeV, may therefore enable one to discriminate between the $b\bar{b}$ and pure leptonic final states.\footnote{Note that $0.1\ \rm GeV$ is actually below the lowest data point for the excess (which is around 0.3--0.4 GeV).} We note that unresolved sources are likely to contribute to the flux in such a small angular region. Hence although the contribution from the leptonic scenario might not be large enough in the inner degree at low energy with respect to observations, the total flux may actually be compatible with the data.

\section{Conclusion}
\label{conclusion}

In this paper, we have demonstrated that taking into account the gamma-ray emission from DM-induced electrons drastically changes the interpretation of the Fermi-LAT excess, since it allows one to obtain an excellent fit to the spectrum of the excess for DM annihilations into leptons only. Therefore, $b\bar{b}$ is not the only viable channel, and we have rehabilitated the pure leptonic channel containing a combination of leptons. More specifically, we have shown that the contributions of the $e^{+}e^{-}$ and $\mu^{+}\mu^{-}$ channels to IC and bremsstrahlung are very important. The reason for this improved fit to the Fermi excess is the IC and bremsstrahlung contributions, which give a gamma-ray spectrum at slightly lower energies than the prompt emission. The effect is strong for democratic annihilation into leptons, while it gets weaker (but definitely non-negligible) for the scenarios favored by the latest constraints \cite{Bringmann_constraints}, with no electrons and a branching ratio into muons of 0.25. Possible additional constraints on this scenario involve the morphology of the gamma-ray flux at low energy: our model is not in strong tension with the morphology of the excess in the energy range of the data, but looking at lower energies may help to discriminate between the leptonic and $b\bar{b}$ scenarios. Therefore, in the absence of such a strong constraint, and should the excess be of DM origin, one would definitely need to take into account these leptonic final states to determine the DM mass and the value of the self-annihilation cross section, even though models may be harder to build than those with a pure $b \bar{b}$ final state \cite{Boehm:2014hva}. 

\acknowledgments{We would like to thank Marco Cirelli, Christopher McCabe, Tim Linden, Lars Bergstr{\"o}m, Christoph Weniger and Torsten Bringmann for fruitful discussions. This research has been supported at IAP by the ERC Project No.~267117 (DARK) hosted by Universit\'e Pierre et Marie Curie (UPMC) - Paris 6 and at JHU by NSF Grant No.~OIA-1124403. This work has been also supported in part by UPMC and STFC.}

\newpage

\bibliographystyle{h-physrev} 
\bibliography{biblio_thesis}

\end{document}